\documentstyle[amssymb,preprint,aps]{revtex}
\tightenlines 

\begin{document}
\title{The one-electron Green's function of the half-filled Hubbard model on a
triangular lattice}
\author{M.C. Refolio, J.M. L\'{o}pez Sancho, and J. Rubio}
\address{Instituto de Matem\'{a}ticas y F\'{i}sica Fundamental, CSIC \\
Serrano 113 bis, 28006 Madrid, Spain\\
e-mail: refolio@imaff.cfmac.csic.es\\
Fax: +34-(1)5854894}
\maketitle
\date{13 2 01}

\begin{abstract}
The one-electron density of states for the half-filled Hubbard model on a
triangular lattice is studied as a function of both temperature and \
Hubbard $U$ using Quantum Monte Carlo. We find three regimes: (1) a
strong-coupling Mott-Hubbard regime, characterized by a gap which persists
even at high temperatures; (2) a weak-coupling paramagnetic regime,
characterized by the absence of a pseudogap at any finite temperature; and
(3) an intermediate-coupling (spiral) spin-density-wave regime,
characterized by a pseudogap which appears when $U$ is increased beyond a
critical (temperature dependent) value. The behavior of the $\sqrt{3}\times 
\sqrt{3}$ adlayer structures on fourth-group semiconductor surfaces is
briefly commented upon in the light of the above discussion.

{\bf Keywords} Hubbard modeling. Semiconducting surfaces. Charge and spin
density waves. Surface Mott insulators.

{\bf PACS:} 73.20.At; 75.30.Fv; 75.30.Pd; 71.15.Mb
\end{abstract}

\section{Introduction}

As is well-known, the frustration effects associated with the triangular
lattice often lead to non-trivial ground-state degeneracies as in the
antiferromagnetic (AF) spin 1/2 Ising model \cite{un,do}. The classical
Heisemberg model on the two-dimensional (2D) triangular lattice with
nearest-neighbor AF coupling and easy-axis exchange anisotropy is another
example where frustration leads to a novel ground-state degeneracy. This
system has attracted much attention especially since Anderson\cite{tres}
suggested the possibility of a resonating-valence-bond ground state for the
spin one-half case. Simply, the quantum liquid of radomly distributed
spin-singlet pairs could be an efficient way to overcome the frustration of
the N\'{e}el state in the triangular antiferromagnet. One more important
source of interest in these lattices is the well-known diversity and
richness of physical phenomena displayed by most transition metal compounds 
\cite{cuatro}.

The experimental observation of low-temperature insulating phases in some $%
\sqrt{3}$-adlayer structures on (111) Si and Ge surfaces has only fostered
the interest in these 2D triangular lattices. Thus, whereas the $\sqrt{3}$%
-overlayers of Sn and Pb on Ge(111) are both metallic at high temperature,
their corresponding low-temperature counterparts are either metallic, as in
the case of Sn\cite{cinco,seis,siete}, or weakly insulating as in the case
of Pb \cite{ocho,nueve,diez}. This latter system seems to go through some
kind of reversible metallic to insulating transition whose precise nature is
still controversial \cite{once,doce,trece}. A charge-density wave (CDW) has
been invoked in the case of Pb (but not in the case of Sn) as the driving
force for the destabilization of the high-temperature phase, a conjecture
not universally accepted at the time of writing. Related isoelectronic
systems, on the other hand, like the $\sqrt{3}\times \sqrt{3}$ adlayer of Si
on SiC(0001) \cite{catorce}, or of K on Si(111):B\cite{quince} show a clear
insulating behavior with a large gap and no phase transitions. These systems
have been studied theoretically both within the local-density (LDA)\cite
{cinco,dieciseis,diecisiete} and the Hartree-Fock\cite{dieciocho,diecinueve}
approximations. Quite recently\cite{veinte} the LDA+U approach has been used
to include strong on-site repulsions in SiC(0001).

In this paper we carry out a model study of the triangular lattice in order
to explore some general questions any realistic theory should comply with.
For instance understanding the temperature behavior of the one-electron
density of states (DOS) is essential to the development of a complete
picture of these metal-insulator transitions. Hence we report the results of
a Quantum Monte Carlo (QMC) simulation of the half-filled Hubbard model on
such a triangular lattice in the grand canonical ensemble. The one-electron
Green's function is studied as a function of both temperature and coupling
constant (Hubbard $U$). As the temperature is lowered, a pseudogap develops
in the one-electron DOS for intermediate values of $U$. This pseudogap is
accompanied by two weak peaks in the spin structure factor which signal the
formation of a spiral spin-density-wave (SDW). For lower $U$, no gap at all
is found even for low temperatures, the system remaining always
paramagnetic. For higher $U$, on the other hand, a well developed gap
appears at any temperature, accompanied by a strong peak in the spin
structure factor. The system is then brought into a state very similar to
the ground state of the triangular antiferromagnet (the three-sublattice
model). We emphasize that these are not distinct phases, but only different
regimes with smooth transitions among them, as characterized by the behavior
of the one-electron DOS. Since the presence or absence of a gap or pseudogap
is of fundamental importance in determining the properties of a system, we
believe that this type of characterization is useful and can be of help in
understanding the electronic properties of the more complex adsorption
systems referred to above.

The paper is organized as follows: Section II recalls some of the basic
properties of the triangular lattice as well as the Hubbard model, in order
to fix the notation. Sec III summarizes very briefly the Quantum Monte Carlo
algorithm in the grand canonical ensemble. The resulting one-electron DOS
and spin structure factor are displayed and discussed in Sec IV and,
finally, the paper closes with some concluding remarks in Sec V

\section{Basic Properties of the Triangular Lattice and Hamiltonian}

In the $\sqrt{3}\times \sqrt{3}$ R30$%
{{}^o}%
$ adlayer structures of Sn or Pb on Ge(111) at one third coverage, each
adsorbate sits on top of a triangle of Ge atoms. With just one unpaired
electron per adsorbate, the overlayer is half-filled and, therefore,
metallic in the absence of electron-electron interactions. This overlayer
can in \ turn be described as a $3\times 3$ lattice of adsorbate triangles
(the three-sublattice model) with three unpaired electrons per triangle and,
therefore, again metallic in the absence of interaction. The corresponding
surface Brillouin zones (SBZ) are the large and small hexagon, respectively,
in Fig 1. The adlayer $\sqrt{3}\times \sqrt{3}$ lattice has just one band in
the large zone, $\varepsilon _{k}^{0}$, which folds onto three bands in the
small zone. Fig 2 shows these three bands, $\varepsilon _{k}^{0},$ $%
\varepsilon _{k}^{+},$ and $\varepsilon _{k}^{-},$ unfolded in the extended
zone scheme in order to see the nesting properties. They are given by $(t$%
=hopping strength) 
\begin{equation}
\varepsilon _{k}^{0}=2t\cos kx+4t\cos \frac{1}{2}k_{x}\cos \frac{\sqrt{3}}{2}%
k_{y}
\end{equation}
\begin{equation}
\epsilon _{k}^{\pm }=-\frac{1}{2}\varepsilon _{k}^{0}\pm t\sqrt{3}\left(
\sin k_{x}-2\sin \frac{1}{2}k_{x}\cos \frac{\sqrt{3}}{2}k_{y}\right)
\end{equation}
It is easy to see that $\epsilon _{k}^{\pm }$ are just $\epsilon _{k}^{0}$\
\ for $k=(k_{x}\pm \frac{2\pi }{3},k_{y})$. These bands cross at the points $%
M$' and $K$'. The wavevector $K$=($4\pi /3,0$) turns out to be a nesting
vector with the band folding around the $M$'=0.5 $K$ point ($M$'$%
K\longrightarrow M$'$\Gamma ,$ band $\varepsilon _{k}^{+})$. Likewise $%
KM\longrightarrow \Gamma M$' (band $\varepsilon _{k}^{-},)$. Fig 3, finally,
shows the resulting band along the small SBZ contour. Notice that the $M$'$K$%
' direction is obtained by folding the $MK$' portion of $\varepsilon
_{k}^{0}.$

This is of no consequence for the interaction-free system at half filling
since the Fermi surface is not anywhere close enough to either the large or
the small SBZ boundaries. When the interaction is turned on, however, the
nesting symmetry may come into play, although weakly, at the points $M$ and $%
M$', closest to the Fermi surface. We shall see that, even at half-filling,
this is indeed the case for the spin structure factor when $U$ is large
enough.

In order to describe the interacting system, we adopt the Hubbard model,
given by the standard Hamiltonian 
\begin{equation}
H=t\sum_{<ij>s}c_{is}^{+}c_{js}-\mu \sum_{is}n_{is}+U\sum_{i}\left(
(n_{i\uparrow }-\frac{1}{2})(n_{i\uparrow }-\frac{1}{2})\right)
\end{equation}
where $t$ is the hopping strength, $U$ the on-site repulsion and $\mu $ the
chemical potential. The single sums run over all the $N\times N$ adlayer
atoms and the symbol 
\mbox{$<$}%
\mbox{$>$}%
means summation over nearest neighbors ($nn$). As usual,\ $c_{is}^{+}$
creates, while $c_{is}$\ destroys, an electron of spin $s$ at site $i$\ with
occupation number $n_{is}=$ $c_{is}^{+}c_{js}.$ We take $t$=0.055 eV so as
to start with a narrow adlayer bandwidth ($W$=$9t)$ of around 0.5 eV at $U$%
=0. U is varied to cover different regimes of the triangular lattice and $%
\mu $ is adjusted so as to have always half filling. Recall that, unlike the
case of bipartite lattices, $\mu =U/2$ does not necessarily correspond to
half-filling since particle-hole symmetry does not hold in a triangular
lattice.

This Hubbard model is now simulated by the QMC approach in the grand
canonical ensemble as initially developed by Blankenbecler et al\cite
{veintiuno} and supplemented by a discrete lattice version of the
Hubbard-Stratonovich transformation by Hirsch\cite{veintidos}. The whole
approach has been explained at length by Hirsch\cite{veintitres} and White
et al\cite{veinticuatro} and is briefly summarized in the following section.

\section{Quantum Monte Carlo Algorithm}

In a grand canonical simulation\cite
{veintiuno,veintidos,veintitres,veinticuatro} the imaginary time is
discretized through the introduction of $L$ time slices separated by an
interval $\Delta \tau $ such that $\beta $=$\Delta \tau L$. The partition
function is then written as 
\begin{equation}
Z=Tre^{-\beta H}=Tre^{-\Delta \tau LH}\eqsim Tr(e^{-\Delta \tau \widehat{K}%
}e^{-\Delta \tau \widehat{V}})^{L},
\end{equation}
where 
\begin{equation}
\widehat{K}=t\sum_{<ij>}c_{is}^{+}c_{js}-\mu
\sum_{is}n_{is}=\sum_{<ijs>}K_{ij}c_{is}^{+}c_{js}
\end{equation}
and 
\begin{equation}
\widehat{V}=U\sum_{i}\left( n_{\uparrow }-\frac{1}{2}\right) \left(
n_{\downarrow }-\frac{1}{2}\right) .
\end{equation}
The last step in Eq(4) follows from the Trotter formula\cite{veinticinco}
which introduces systematic errors in the measured quantities of order $%
\Delta \tau ^{2}$. One should therefore like to take $\Delta \tau $ as small
as possible, although keeping the number of time slices not too large. This
poses a serious limitation at low temperatures (large $\beta ).$

A discrete Hubbard-Stratonovich (HS) transformation\cite{veintidos} is now
performed for each on-site interaction term within any time slice. 
\begin{equation}
e^{-\Delta \tau U\left( n_{i\uparrow }-\frac{1}{2}\right) \left(
n_{i\downarrow }-\frac{1}{2}\right) }=\frac{1}{2}e^{-\Delta \tau
U/4}Tr_{\sigma }e^{-\Delta \tau \lambda \sigma _{il}(n_{i\uparrow
}-n_{i\downarrow })}
\end{equation}
where a discrete, auxiliary Ising field $\sigma _{il}=\pm 1$ ($l$ runs
through the $L$ time slices) has been introduced and $\cosh \lambda \Delta
\tau =e^{U\Delta \tau /2}$. $Tr_{\sigma }$ means summation over the $NL$
Ising ''spins'' $\sigma _{il},$ which define the auxiliary Ising field
through the lattice sites and time slices. Defining now $N\times N$ matrices 
$V(l)$ such that 
\begin{equation}
V_{ij}(l)=\frac{1}{2}\delta _{ij}\lambda s_{ij}\text{,}
\end{equation}
we can write 
\begin{equation}
Z=\left( \frac{1}{2}e^{-\Delta \tau U/4}\right) ^{NL}Tr_{\sigma }Tr\Pi
_{s,l}D_{l}^{s},
\end{equation}
where 
\begin{equation}
D_{l}^{s}=e^{-\Delta \tau \sum_{ijs}K_{ij}c_{is}^{+}c_{js}}e^{-\Delta \tau
\sum_{is}V_{ii}(l)\alpha _{s}n_{is}}
\end{equation}
where $\alpha _{s}$=$\pm 1$ \ for $s$=$\uparrow \downarrow $ respectively.
Notice carefully the difference between $\sigma $, Ising spin, and $s$,
fermion spin.

The fermion degrees of freedom can be traced over\cite{veintiuno,veintitres}
to yield 
\begin{equation}
Z=Tr_{\sigma }\Pi _{s}\det 0_{s}\left[ \sigma \right]
\end{equation}
where $0_{s}\left[ \sigma \right] ,$\ the fermion matrix, is given by 
\begin{equation}
0_{s}=I+B_{l}^{s}B_{l-1}^{s}...B_{1}^{s}
\end{equation}
with 
\begin{equation}
B_{l}^{s}=e^{-\Delta \tau K}e^{-\Delta \tau \alpha _{s}V(l)}
\end{equation}
and $I$ \ being the unity matrix. Therefore $\Pi _{s}\det 0_{s}$ plays the
role of a Botzmann weight for evaluating the average of any operator, $<A>$=$%
TrAe^{-\beta H}/Tre^{-\beta H}$. The above steps must now be repeated for $%
TrAe^{-\beta H}$\ .Thus for $A$=$c_{is}c_{js}^{+}$, the resulting expression
is 
\begin{equation}
Tr\left( c_{is}c_{js}^{+}e^{-\beta H}\right) =Tr_{\sigma }\left(
0_{s}^{-1}\right) _{ij}\Pi _{s}\det 0_{s}\left[ \sigma \right]
\end{equation}
and hence an equal-time single particle Green's function $<c_{is}c_{js}^{+}>$
is obtained by averaging the corresponding matrix elements of the inverse of
the fermion matrix $0_{s}$. Time dependent Green's functions can also be
calculated and involve slightly more complicated matrices\cite{veintitres}.
For any HS field configuration, Wick's theorem for operator products applies
and one can therefore easily calculate time-dependent correlation functions
in terms of single-particle Green's functions for any given Ising
configuration. These imaginary-time quantities are then to be analytically
continued to real time and frequencies via the maximum-entropy method\cite
{veintiseis,veintisiete} which yields the corresponding spectral weight
functions.

The heat-bath algorithm is used to sample the HS\ field subject to the
Boltzmann weight given above. This sampling is normally accomplished by
single spin-flips $\sigma _{il}\longrightarrow $-$\sigma _{il}$, the
corresponding weight-ratio being related to equal-time Green's function\cite
{veintiuno,veinticuatro}. If the new configuration is accepted, the
corresponding Green's functions is updated through simple operations\cite
{veintiuno,veinticuatro}. Calculating the weight ratio requires $N^{2}$
operations, and therefore $N^{3}L$ operations must be done for a sweep
through the whole Hs field. Normally two hundred warm-up sweeps and one
thousand measurements separated by two sweeps were performed for each set of
parameters. At low temperatures the algorithm becomes unstable due to the
large number of time slices required in order to make $\Delta \tau $\ small
enough. The product of B matrices in Eq(12) is ill-conditioned and one has
to resort to matrix factorization techniques\cite{veinticuatro}.

\section{The Three Coupling Regimes}

According to the Mermin-Wagner theorem\cite{veintiocho}, infinite-range
magnetic order is forbidden in two dimensions at any $T\neq 0$. This is so
because the Goldstone modes strongly disorder the system giving rise to a
spin-spin correlation length which decays with temperature as $\xi
(T)\thicksim \exp (A/T),$ where $A$ is a constant. Thus no phase transitions
of magnetic origin can take place in an infinite system except, perhaps, at $%
T$=$0$. Other kinds of phase transitions are outside the scope of this
theorem. Such is the case, e.g., of the (Mott) paramagnetic metal -
paramagnetic insulator transition. Let us specialize to the case of the
triangular lattice.

\subsection{The one-electron DOS}

The one-electron DOS is given by 
\begin{equation}
N(\omega )=\frac{1}{N}\sum_{k}A(k,\omega )
\end{equation}
where $N$ is the number of lattice sites and $A(k,\omega )$, the
spectral-weight function, is the imaginary part of the retarded one-electron
Green's function.

Mean-field studies at $T$=$0$\cite{veintinueve} have\ shown that the
half-filled triangular lattice is a paramagnetic metal in the weak-coupling
regime, in contrast with the SDW insulating character of the square lattice
for small $U/t$. No gap in the one-electron DOS is, therefore, expected at
any temperature for an $infinite$ triangular lattice. It has been shown,
however, that size effects are very strong in this regime \cite{treinta}. A
gap in the one-electron DOS develops as soon as the spin-spin correlations
extend over the whole system. Thus, for lattices of increasing size $N\times
\ N$, the system evolves from a situation where the correlation length $\xi
(T)>N$, (with a gap) to one where $\xi (T)<$ $N$ (without a gap). One should
be careful when drawing conclusions about the existence of gaps from small
lattices.

Fig 4 shows the one-electron DOS of a half-filled $4\times 4$ triangular
lattice with periodic boundary conditions. We have taken $U/t$ =$5$ ($%
\thicksim $ half the bandwidth, $9t$), which is a weak to moderate value,
and several values of $\beta $, $\beta t$=$5$, $10$, $15$ and $20.$ Even for 
$\beta t$\ as high as $20$, the system is far from having a fully developed
gap. Since for the bigger lattices one expects weaker pseudogaps, it may be
safely concluded that in the weak coupling regime a triangular lattice has
no gaps at any temperature, in accordance with the Mermin-Wagner theorem.

In the strong-coupling regime at $T$=$0$ the system is brought into a
commensurate, three-sublattice, $120%
{{}^o}%
$ twist SDW state (similar to the ground-state of the classical
antiferromagnet) which is insulating and stable for increasing $U$. Quantum
fluctuations about the classical antiferromagnetic solution lead to the
essential qualitative physics of the Mott-Hubbard insulator at finite
temperatures with a charge gap of order $U$ in the spectral-weight function.
Fig 5 bears the same information as Fig 4, but with $U/t$=$20$ which is deep
inside the strong-coupling regime. Since size effects are very small in this
regime\cite{treintaiuno}, it is fairly clear that a fully developed gap is
present at any temperature.

We thus see that, for a given temperature, the system evolves from a gapless
situation at small $U$ to a fully developed gap at large $U$. As $U$
increases through the intermediate-coupling regime, one should find a
critical value $U_{c}(T)$ for which the gap first appears. Fig 6 displays,
as Fig 4 and 5, the one-electron DOS for an intermediate value of $U/t$=$10$
($\thicksim $ the bandwidth). As the temperature is lowered from $\beta t$= $%
5$ down to $\beta t$= $20$, an incipient pseudogap gradually evolves into a
fully developed gap. This value of $U$ is clearly below the critical $U$ for
all $\beta t$%
\mbox{$<$}%
$20$, i.e., $U_{c}$ =$10t$ for $\beta t$= $20$. The complementary view is
given in Fig 7, which shows the one-electron DOS for $\beta t$= $5$ and $U/t$%
=$5,10,15$, and $20$. We see the system evolving from a gapless regime to a
pseudogap, a deep pseudogap and finally a fully developed gap. Thus $U_{c}$=$%
20t$ for $\beta t$= $5.$ In this way one generates a temperature-dependent
critical value of the coupling constant $U_{c}(T).$

\subsection{ \ \ The spin structure factor}

The spin structure factor, $s(k)$, is given by the $k$-Fourier transform of
the static spin-spin correlation function 
\begin{equation}
s_{ij}=<\sigma _{iz}(\tau ^{+})\sigma _{jz}(\tau )>_{\tau =0}
\end{equation}
where $\tau ^{+}$=$\tau +o^{+}$ in the imaginary-time domain, and $\sigma
_{iz}$=$n_{i\uparrow }-n_{i\downarrow }$.

The peaks of $s(k)$ and corresponding widths in $k$-space give information,
as is well-known, about the SDW's sustained by the system and the spin-spin
correlation length. Fig 8 shows $s(k)$ for $\beta t$= $20$\ and $U/t$=$5,10$%
,and $20$, representative values of the three regimes. The almost
featureless shape for $U/t$=$5$ evolves into two sharp, although small,
peaks close to $M$ and $M$' in the intermediate regime and, finally into a
large peak at $M$' in the strong-coupling regime. The system,
correspondingly, evolves from a paramagnetic metal, through an
incommensurate spiral SDW, into a Mott-Hubbard insulator

\section{Concluding Remarks}

The variation of the one-electron DOS with both temperature and coupling
constant seems a useful tool for the purpose of identifying the different
regimes of a given system. For the special case of the half-filled repulsive
Hubbard model on a triangular lattice, we have identified an intermediate,
temperature dependent coupling regime which interpolates smoothly between
the weak-coupling (paramagnetic metal) and the strong-coupling (Mott-Hubbard
insulator) regimes. As the temperature is lowered in this
intermediate-coupling regime, the system evolves from metallic to insulating.

We conclude with a comment on the $\sqrt{3}\times \sqrt{3}$\ adlayer
structures on group fourth semiconductor surfaces. Although a close
connection with the above model study is not claimed, these structures may
constitute a physical realization of the three coupling regimes just
described, Sn/Ge, Pb/Ge, and SiC being examples of the weak, intermediate
and strong coupling regimes, respectively. Despite the added complexity due
to electron-phonon interactions and atomic relaxation of both adsorbate and
substrate atoms, the model study carried out here provides a general
framework for the study of those systems.

\bigskip {\bf Acknowledgments}. This work was supported by the DGICYT
(Spain) Project N%
${{}^o}$%
PB98-0683.

\newpage

\section{\protect\bigskip Figure Captions}

\begin{quote}
{\bf Fig 1. }Large (outer exagon) and small (inner exagon) surface Brillouin
zones (SBZ) of the triangular lattice. Shown are the especial points $\Gamma
,M^{\prime },K,M,$ and $K$' which delimit the contours $\Gamma KM\Gamma $and
\ $\Gamma M^{\prime }K^{\prime }\Gamma $ used in the text.

{\bf Fig 2}. Band structure of the triangular lattice for $U$=0 in the
three-sublattice model. The three bands, $\varepsilon _{k}^{0}$ (main band)
and $\varepsilon _{k}^{\pm },$\ are displayed along the $\Gamma KM\Gamma $\
\ \ \ \ \ \ \ \ contour of the large SBZ in order to show the band crossings
and nesting symmetry.

{\bf Fig 3}. Same as Fig 2, but along the $\Gamma M^{\prime }K^{\prime
}\Gamma $\ contour of the small SBZ.

{\bf Fig 4}. One-electron density of states (DOS) of the triangular lattice
for $U/t$=$5$ (weak coupling) and decreasing temperature, \ $\beta t$=$%
5,10,15$, and $20$

{\bf Fig 5}. Same as Fig 4, but for $U/t$=$20$ (strong coupling).

{\bf Fig 6}. Same as Figs 4 and 5, but for $U/t$=$10$ (intermediate
coupling).

{\bf Fig 7}. One-electron DOS of the triangular lattice for increasing $U/t$=%
$5,10,15$, and $20$ at a fixed temperature \ $\beta t$=$5$.

{\bf Fig 8}. Low-temperature ($\beta t$=$20)$ spin structure factor, $s(k)$,
of the triangular lattice in the weak ($U/t$=$5$\ ), intermediate ($U/t$=$%
10) $ and strong ($U/t$=$20$\ ) coupling regimes
\end{quote}


\bigskip

\bigskip

\begin{references}
\bibitem{un}  \ \ G.W. Wannier, Phys Rev {\bf 79}, 357 (1950)

\bibitem{do}  \ \ J. Stephenson, J.Math. Phys {\bf 5} 1009 (1964)

\bibitem{tres}  \ \ P.W. Anderson, Mater. Res. Bull, {\bf 8}, 153 (1973)

\bibitem{cuatro}  H.F. Pen, J.van den Brink, D.I. Khomskii, and G.A.
Sawatzky, Phys. Rev. Lett, {\bf 78}, 1323 (1997)

\bibitem{cinco}  \ \ J. M. Campinelli, H.H. Weitering, E.W. Plummer and R.
Stumpf, Nature 381, 398 (1996)

\bibitem{seis}  \ \ A. Mascaraque, J. Avila, E.G. Michel and M.C. Asensio,
Phys. Rev. B. {\bf 57} 14758 (1998)

\bibitem{siete}  \ \ A. Goldoni, C. Cepek, and S. Modesti, Phys. Rev. B {\bf %
55}, 4109 (1997).

\bibitem{ocho}  \ \ J. M. Campinelli, H.H. Weitering, M. Bartkowiak, and
E.W. Plummer, Phy. Rev. Lett.{\bf 79, }2859 (1997).

\bibitem{nueve}  A. Goldoni, and S. Modesti, Phys. Rev.Lett. {\bf 79}, 3266
(1997).

\bibitem{diez}  \ G. Lelay, V.Y. Aristov, O. Bostr\"{o}m, J.M. Layet, M.C.
Asensio, J. Avila, V. Huttel, and A.Cricenti, Appl. Surf. Sci.{\bf 123-124},
440, (1998)

\bibitem{once}  \ A. Mascaraque, J. Avila, J. Alvarez, M.C. Asensio, S.
Ferrer, and E.G. Michel, Phys. Rev. Lett. {\bf 82}, 2524 (1999).

\bibitem{doce}  \ J. Avila, A. Mascaraque, E.G. Michel, M.C. Asensio, J.
Ortega, R. Perez, and F. Flores, Phys. Rev. Lett. {\bf 82}, 442 (1999)

\bibitem{trece}  \ H.H. Weitering, J. M. Campinelli, A.V. Malechko, J.
Zhang, M. Bartkowiak, and E.W. Plummer, Science {\bf 285}, 2107 (1999).

\bibitem{catorce}  L.I. Johansson, F. Owman, and P. Martensson, Surf. Sci. 
{\bf 360}, L478 (1998);

J.M. Tremlim, I. Forbeaux, V. Langlais, H. Belkir, and J.M. Debever,
Europhys. Lett. {\bf 39}, 61 (1997)

\bibitem{quince}  H.H. Weitering, X. Shi, P.D. Johnson, J. Chen, N.J.
Dinardo, and S. Kempa, Phys. Rev. Lett. {\bf 78}, 1331 (1997).

\bibitem{dieciseis}  K.W\"{u}rde, P. Kr\"{u}ger, A. Mazur and J. Pollman,
Surf. Rev. and Lett. {\bf 5}, 105 (1998)

\bibitem{diecisiete}  S. Scandolo, F. Ancilotto, G.I. Chiarotti, G. Santoro,
S. Serra and E. Tosatti, Surf. Sci. 402-404, 808 (1998)

\bibitem{dieciocho}  G. Santoro, S. Sorella, F. Becca, S. Scandolo and E.
Tosatti, Surf. Sci. 402-404, 802 (1998)

\bibitem{diecinueve}  G. Santoro, S. Scandolo and E. Tosatti, Phys. Rev B 
{\bf 59}, 1891 (1999)

\bibitem{veinte}  V.I. Anisimov, A.F. Bedin, M.A. Korotin, G. Santoro, S.
Scandolo and E. Tosatti, Phys. Rev B {\bf 61}, 1752 (2000)

\bibitem{veintiuno}  R. Blankenbecker, D.J. Scalapino, and R.L. Sugar, Phys
Rev. D {\bf 24}, 2278 (1981)

\bibitem{veintidos}  J.E. Hisch, Phys. Rev B {\bf 28}, 4059 (1983)

\bibitem{veintitres}  J.E. Hisch, Phys. Rev B {\bf 31}, 4403 (1985)

\bibitem{veinticuatro}  S.R. White, D.J. Scalapino, R.L. Sugar, E.Y. Loh,
J.E. Gubernatis, and R.T. Scalettar, Phys. Rev B {\bf 40}, 506 (1989)

\bibitem{veinticinco}  M. Suzuki, Prog Theor.Phys. {\bf 56}, 1454 (1976)

\bibitem{veintiseis}  R.N. Silver, D.S. Sivia, and J.E. Gubernatis, Phys.
Rev B {\bf 41, }2380 (1990)

\bibitem{veintisiete}  R.N. Silver, J.E. Gubernatis, D.S. Sivia and M.
Jarrell, Phys. Rev. Lett. {\bf 65, }496 (1990)

\bibitem{veintiocho}  N.D. Mermin and H. Wagner, Phys Rev. Lett {\bf 17}
1133 (1966)

\bibitem{veintinueve}  H.R. Krishnamurthy, C. Jayaprakash, Sanjoy Sarker,
and Wolfgand Wenzel, \ Phys Rev. Lett. {\bf 64}, 950 (1990)

\bibitem{treinta}  S.R. White, Phys Rev B {\bf 46}, 5678 (1992)

\bibitem{treintaiuno}  S.R. White, Phys. Rev. B {\bf 47}, 1160 (1993)
\end{references}
\end{document}